\documentclass[letter]{aa}
\usepackage{graphicx}
\usepackage{txfonts}
\usepackage{natbib}
\usepackage{graphicx}

\bibpunct{(}{)}{;}{a}{}{,} 

\begin{document}

\title{Properties of the short period CoRoT-planet population I:}

\subtitle{Theoretical planetary mass spectra for a population of stars\\
   of 0.8 to 2 solar masses and orbital periods of less then 20 days.}

\author{G. Wuchterl \and C. Broeg \and S. Krause \inst{1}
\and B. Pe\v{c}nik \inst{2} \and J. Sch{\"o}nke \inst{3}
       }

\offprints{G. Wuchterl}


   \institute{Th\"{u}ringer Landessternwarte Tautenburg, Sternwarte 5
                    , D--07778 Tautenburg, Germany
                \and Department of Physics, University of Split, N. Tesle 12
                    , 21000 Split, Croatia
                \and Institut f{\"u}r Theoretische Physik, Universit{\"a}t Bremen
                   , Otto-Hahn-Allee 1, D--28359 Bremen, Germany
            }

\date{Received December, 27th 2006/ accepted ?}

\abstract
  {
  We study the planet populations in the discovery window of the CoRoT-space-telescope
  scheduled for launch on December 27th. We base the prediction on `first principles'
  calculations of planet formation in the framework of the planetesimal hypothesis.
  }
  {
  To provide a-priori planetary initial mass functions for confrontation with
  the CoRoT-planet discoveries in the entire range of sensitivity of the CoRoT
  instrument, i.e.\ for all giant planets and down to terrestrial planet
  masses.
  }
  {
  We construct a comprehensive set of static complete-equilibrium core-envelope
  protoplanets with detailed equations of state and opacity and radiative transfer
  by convection and radiation. Protoplanets are calculated for host-star masses of
  0.8 to 2 solar masses and orbital periods of 1 to 16 days. We subsequently check
  the stability of the planetary population by a series of methods.}
  {
  We find the static planetary populations to be stable and thus a
  plausible ensemble to predict the planetary IMF for orbital periods in the specified
  range.}
  {
  We predict bimodal planetary initial mass functions with shapes depending on orbital
  period. The two main maxima are around a Jupiter mass and about 50 earth masses. We
  predict an abundant population of Hot Neptunes and a large population of
  planets that fill the solar-system gap of planetary masses between Neptune
  and Saturn.
  }


\maketitle

  \section{Introduction}

  Planets form in circumstellar disks which are a necessary step in star-formation.
  Presently two mechanisms are in discussion:
  formation via a disk-instability and via planetesimals, see
  \citet{2000prpl.conf.1081W}. Uncertainties about the actual nebula conditions
  at the era of planet formation and an incomplete understanding of relevant
  physical processes (e.g.\ of planetesimal formation and of the various types
  of  orbital migration) limit the predictive power of conventional formation
  theories. In preparation for the CoRoT-transit-search for exoplanets, e.g.\
  \citet{2002sshp.conf...17B} we have developed a new synoptic approach to planet
  formation, (\citet{2003DiplPecnik,2005A&A...440.1183P,2006DissBroeg}).

  \section{Synopsis of protoplanets}

  \citet{2006DissBroeg} calculated all planets and protoplanets in hydrostatic
  and thermal equilibrium for stellar masses of 0.4, 0.8 , 1 and
  $2\,{\rm M_{\sun}}$\footnote{The stellar types that can be observed with high
    photometric precision and are likely to be found in CoRoT's fields. The
    period range results from the 150d continuous observation periods and
    the requirement to observe three transits to make a safe detection.}
   and orbital periods of 1, 4, 16 and 64 d (0.02, 0.049, 0.124 and
  0.313 AU for the solar case). The ranges are chosen to cover CoRoT's
  discovery space.
  Planetary models are spherically symmetric, consist of a rigid, constant
  density core and a gaseous envelope in hydrostatic and thermal equilibrium.
  Planetesimal accretion provides an energy input at the core surface that
  is transferred through the envelope by radiation and convection. Detailed
  equations of state and opacities for roughly solar composition are used
  with 'interstellar' dust opacities. Orbital distance enters via the thermal
  and tidal effects of the star. Assuming that no core size is favoured and
  that gravitationally stable nebulae provide the pressures for embedding
  the planets, statistical properties of the planet population are provided,
  in particular planetary mass spectra. Here we use the \emph{CoRoT~Mark~1~v1.1}
  ensemble of planetary models, \cite{2006DissBroeg}, \emph{Mark~1} for short.
  Mass spectra and solution manifolds are available online, \citet{2006Mark_1_v1.1_online}, see
  \citet{2006tafp.conf...70B} for  an application to the
  large-core Pegasi-planet HD 149 026 b.
  Here we apply the method to theoretically predict what planets are
  in CoRoT's search space. We restrict the discussion to close-in planets
  with orbital periods of $<16\,{\rm d}$ for two reasons: (1) they are in CoRoT's
  discovery window for the first run with a foreseen continuous observing
  period of 30--60~d and hence with a completeness limit for planets
  detected by at least three transits at about 15 days, (2) there is a
  qualitative change in the planet-population for periods of $\ga 16\,{\rm d}$,
  \citet{2006DissBroeg} and planet formation being dominated by hydrostatic
  processes for smaller periods, \citet{2006tafp.conf...70B}.

  \section{Planetary Mass Spectra for a typical CoRoT field}

\begin{figure*}[htbp]
  \begin{center}
  \begin{tabular}{p{5.7cm}p{5.7cm}p{5.7cm}}
    \includegraphics[width=5.5cm]{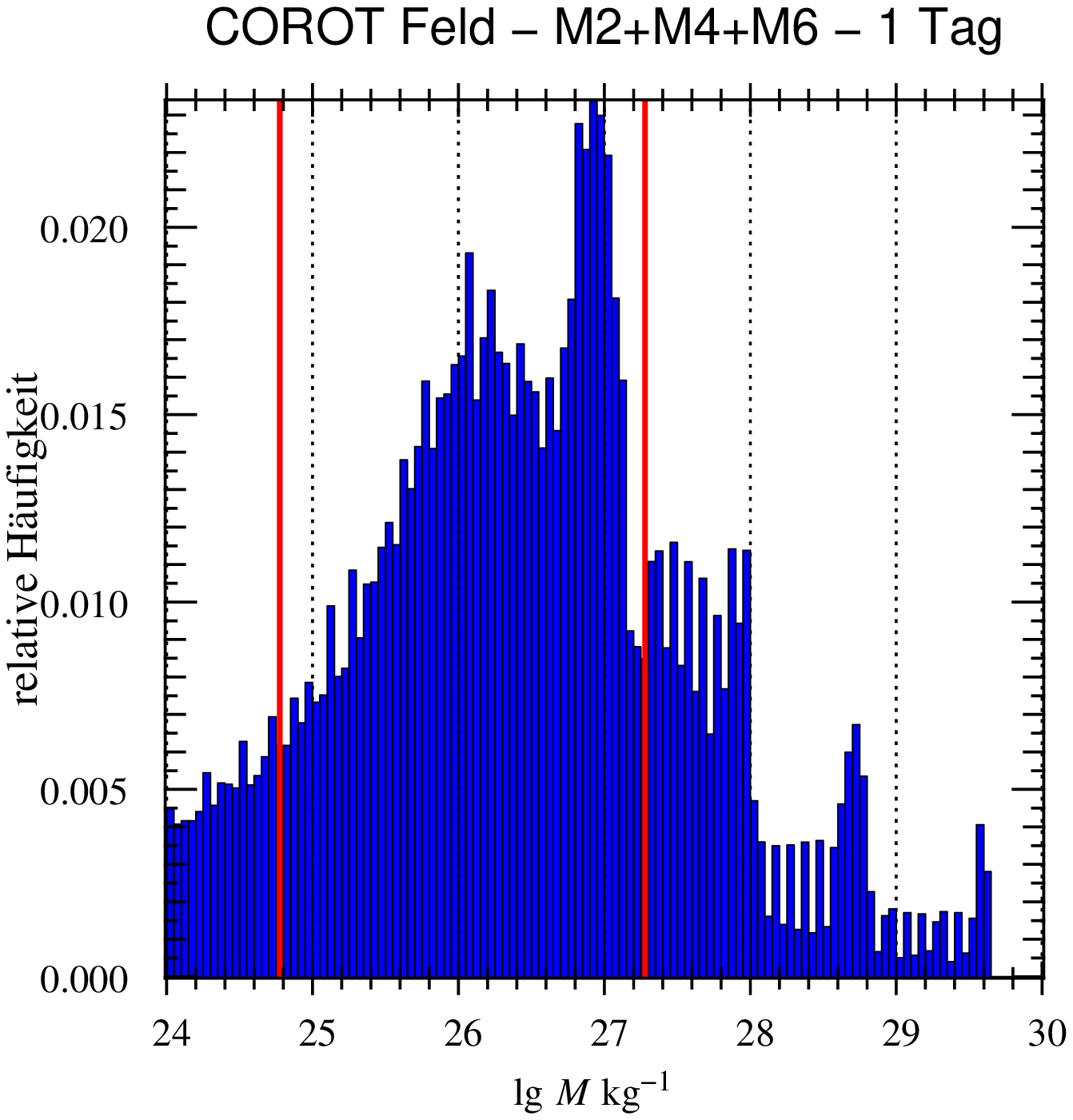}
    &
    \includegraphics[width=5.5cm]{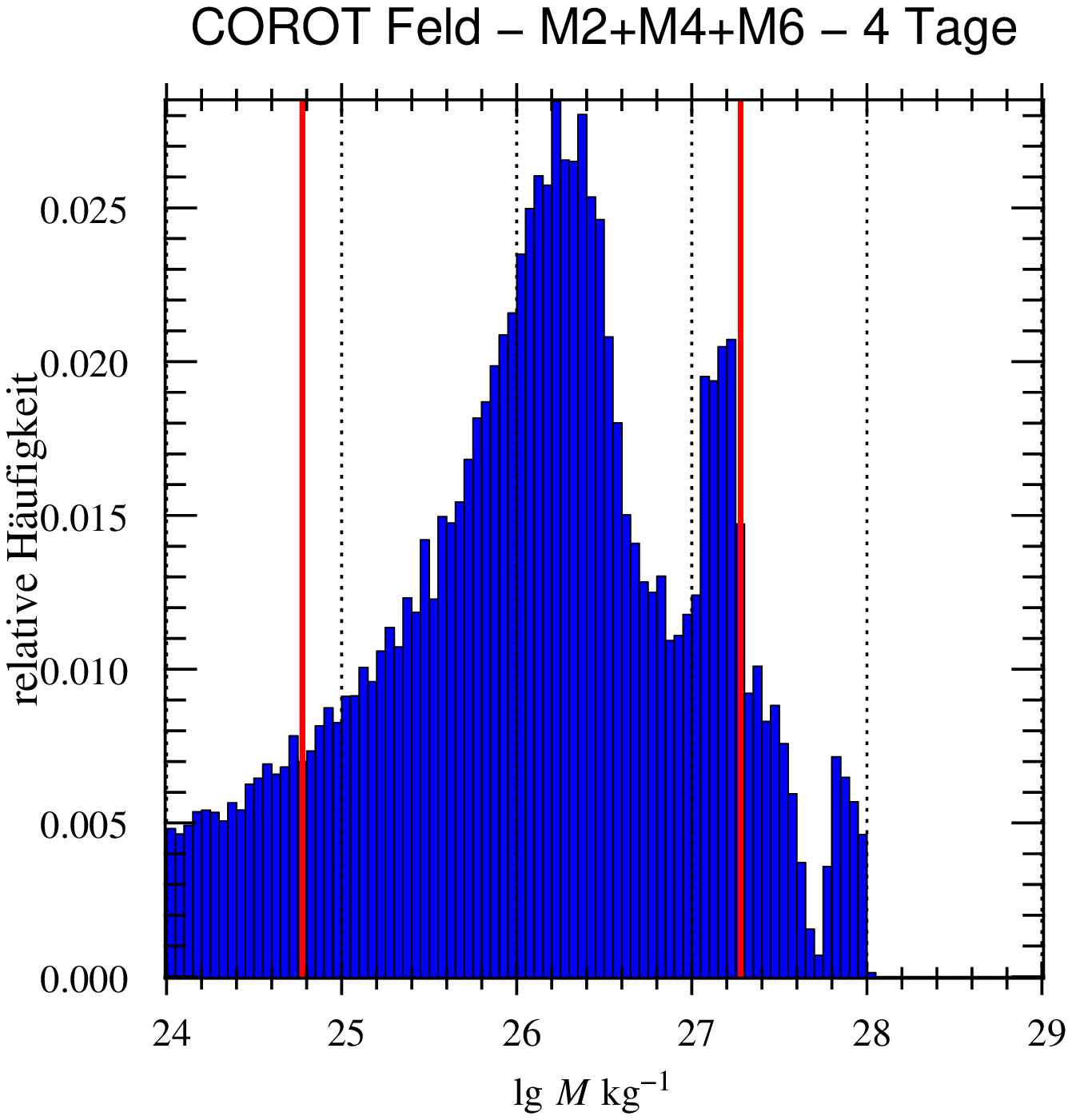}
    &
    \includegraphics[width=5.5cm]{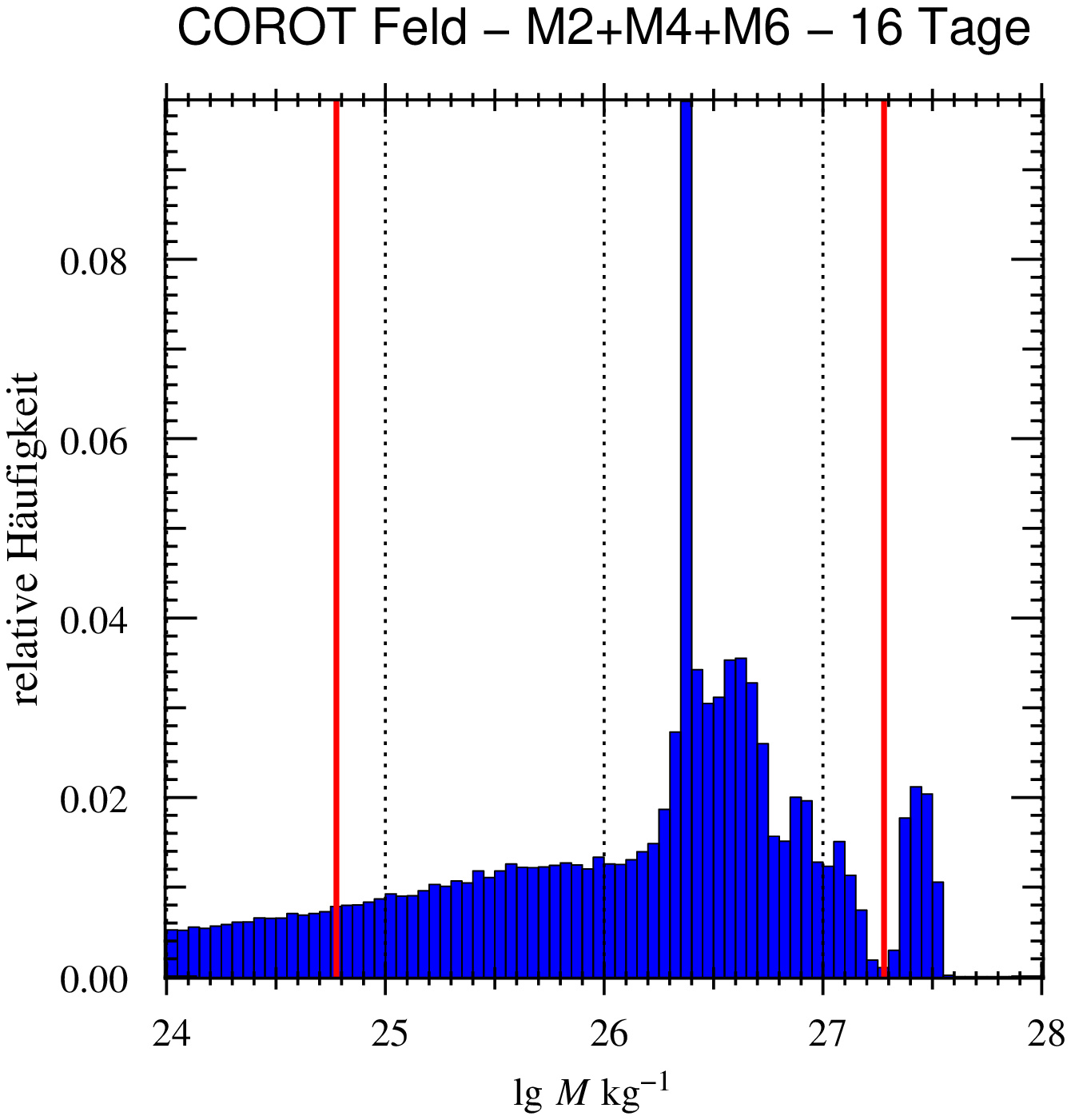}
    \end{tabular}
    \end{center}
    \caption{Theoretical planetary initial mass functions calculated from
        planet formation theory for a typical CoRoT-field. Results are shown
        for planetary orbital periods of 1, 4, and 16 days, from left to right.
        The relative frequency is plotted as function of lg
        mass in kg. Vertical red lines mark the Earth and Jupiter masses.
        $\sim 10^6$ planetary models in total. Structures of width
        $< 0.3$ dex have to be taken with care, because of undersampling
        in spectral type due to the unexpected richness of the mass-spectra.
        (`M2+M4+M6' designates planetary core-accretion and is not related
        to the stellar population).
            }
    \label{Mark1_CoRoT_Mix}
 \end{figure*}

  Our calculations of the planet population show a strong dependence
  of the planetary properties on stellar mass, \citet{2006DissBroeg}.
  We therefore account for the types of stars CoRoT will observe.
  Given the Mark~1 stellar mass-grid, we model a typical CoRoT-field with
  a mixture of 56 \%, 37 \%, 7\% of 2, 1, and $0.8\,{\rm M_{\sun}}$ stars, resp..
  This roughly represents a grouping of stellar spectral types
  A + F, F + G and K, resp.. 
  As noted by \citet{2006DissBroeg} Mark~1 has a relatively sparse
  sampling of the  stellar mass-dependence but is already based on $\approx 10^6$
  planetary models und thus by far the largest survey of planet-formation.
  We do not include the Mark~1 results for
  $0.4\,{\rm M_{\sun}}$ (M-stars) here. To account for the uncertainties
  in planetesimal-formation and -accretion theory we use the results \emph{for
  all core-accretion rates} given in Mark~1,
  i.e.\ $10^{-6}$, $10^{-4}$, and $10^{-2}\,{\rm M_{Earth} a^{-1}}$. The
  resulting theoretical planetary mass spectra for a CoRoT-stellar
  population mix, the \emph{CoRoT-Mix} for short, are shown in Fig.~\ref{Mark1_CoRoT_Mix}.
 We note that this a-priori prediction of the planetary initial mass function
 naturally predicts known empirical `planetary' masses. Given the fact that the
 only particular macroscopic input masses are the stellar masses, that is
 non-trivial. For a general discussion of the properties we refer to
 \citet{2006DissBroeg} and focus on the particularities of the CoRoT-Mix in
 the following.

 The multi-peak structure in the mass functions follows the
 general double peaked structure described by \citet{2006DissBroeg} but is
 complicated by mixing the mass spectra of different host stars. This results
 is the structures displayed in Fig.~\ref{Mark1_CoRoT_Mix}. Averaging the mass spectra using a finer sampling
 in host star mass would most likely revert the mass spectra to a double peaked
 structure.
 At all periods the most frequent planets appear with the first peak near
 Jupiter's mass and the second one close to Neptune's. Both peaks move to larger masses with increasing
 period or orbital distance. The shift is about half an order of magnitude.
 The peaks in the mass spectra become narrower for larger distance indicating
 that the typical planetary masses (Neptune and Jupiter) become more well
 defined farther from the host stars. A key feature at all periods is a
 rich population of `Hot Neptunes' that always significantly outnumbers the
 Jupiters.

 For reference and an accompanying study of the impact of particle mass-loss
 processes characteristic numbers are given for the pure solar mass results of
 CoRoT-Mark~1 in Tab.~1.
   \begin{table}

%
     $$
     \begin{array}{rrrrrrr}
     \hline
     \noalign{\smallskip}
     T_{\rm orb}/ {[\mathrm{d}]} &  a / {[\mathrm{AU}]} &  T_{\rm neb} / {[\mathrm{K}]}
        & M_1 / {[\mathrm{kg}]}^{\mathrm{a}} & M_2 / {[\mathrm{kg}]^{\mathrm{a}}}
        & P_1 / {[\mathrm{Pa}]}^{\mathrm{a}} & P_1 / {[\mathrm{Pa}]^{\mathrm{a}}} \\
     \noalign{\smallskip}
     \hline
     \noalign{\smallskip}
      1 & 0.02 & 2001.5 & 1.0\,10^{26} & 1.0\,10^{27} & 1.0\,10^{4} & 1.0\,10^{5}  \\
      4 & 0.05 & 1260.9 & 2.5\,10^{26} & 1.6\,10^{27} & 1.0\,10^{4} & 1.0\,10^{2}  \\
     16 & 0.13 &  794.3 & 3.2\,10^{26} & 3.2\,10^{27} & 1.0\,10^{4} & 1.0\,10^{-2} \\
     64 & 0.30 &  500.4 & 3.2\,10^{26} & 1.0\,10^{27} & 1.0\,10^{4} & 1.0\,10^{4}  \\
     \noalign{\smallskip}
     \hline
     \end{array}
     $$
\begin{list}{}{}
\item[$^{\mathrm{a}}$] Estimated from \citet{2006DissBroeg}, Figs.~8.2, B.3 and
B.11.
\end{list}
\caption[]{Characteristic quantities of the CoRoT-Mark~1~v1.1 theoretical
        planetary population for a solar mass host star, from \citet{2006DissBroeg}
        for various orbital periods, $T_{\rm orb}$. Semi-mayor axis, $a$ and the
        radiative equilibrium nebula-temperature, $T_{\rm neb}$ result from the host
        star mass and luminosity, resp.. $M_1$ and $M_2$ give characteristic
        values for peaks in the planetary mass spectra. $P_1$ and $P_2$ are the
        corresponding typical nebula pressures.}
         \label{Mark1SoMa}
\end{table}

\section{Stability of the planetary population}

The CoRoT~Mark~1 planet population is based on an equilibrium concept, much
like the stellar main sequence where stars are in complete i.e.\ hydrostatic
and thermal equilibrium. The advantage is that whatever the formation history
of the planet and the nebula was, it does not influence the planetary
properties. Thus whether planets migrated to their final position or had
undergone giant impacts or mass-loss events during the formation era does not
change our prediction as long as the respective conditions of the planetary
equilibria are realised --- i.e. nebulae are sufficiently diverse --- as the
planets approach their final masses. The disadvantages are (1) that because the
method is based on time-independent models, we do not automatically have
information about the stability of the equilibria used in the construction and
(2) because it is based on planet formation theory we do not account for
modifications of the planetary properties during planetary evolution in to the
present, in particular and most important mass loss, that might distort the
mass spectra. The second issue is discussed in an accompanying letter, the
question of stability in the following sections. For HD 149 026 b,
\citet{2006tafp.conf...70B} 
showed that planet formation is a completely hydrostatic process in that case.
They gave a complete formation-history including a radiation fluid-dynamical
calculation with convection and detailed, particle in box planetesimal
accretion to ages of about 100 Ma. Because that is presently beyond
computational feasibility for the entire Mark~1 population of $\sim 10^6$
planets we have addressed the stability question by a series of simplified
model systems: (1) linear analysis of isothermal and polytropic protoplanets,
(2) non-linear fluid dynamics for isothermal protoplanets, (3)
quasi-hydrostatic stellar evolution type calculations to investigate non-linear
secular or thermal stability and (4) radiation-fluid dynamical calculations
with convection.

\section{Connections between mass spectra and linear dynamical stability analysis}

A comprehensive linear analysis of isothermal protoplanets was given by
\citet{2005DiplSchoenke} who found two types of linear-instabilities that
operate in protoplanets for sufficiently high-nebula-densities and are damped
by large cores. One characteristic peak in the mass spectra obviously results
from structures in the outer envelope, which are in a state describable as the
onset of self-gravity. These models were found for a certain region in the
parameter set of core mass and gas pressure at the core surface (cf. `region
IV´  in \citet{2005A&A...440.1183P} or \citet{2006DissBroeg}) and have all the
same characteristic mass, leading to the peak in the spectra. The linear
stability analysis (cf.\ \citet{2005DiplSchoenke}) shows a clear connection
between this class of models and a dynamical instability. Almost all models are
linearly unstable, except for a small strip in the low pressure regime. What
kind of consequences for the spectra follow from this result can not be
answered completely from the standpoint of the linear analysis because no
information about the dynamical evolution can be gathered from it. It is likely
that a part of the unstable models will end up in the stable low pressure
strip, a behaviour which the non-linear stability analysis shows (cf.\
\citet{pecnik05}, \citet{PecnikWuchterl_06}). At least we can say that a
`stability-corrected' mass spectrum would reduce the `strength' of the peak
significantly. The appearance of the described unstable models mainly depends
on the available envelope volume (i.e. the Hill radius) and therefore on the
orbital distance or period. As the Hill radius becomes small (i.e.\ orbital
distance or period are small), the region where unstable envelope structures
are built up shifts to smaller core masses and the spectral peak broadens.
Because of the quantitative differences between isothermal and detailed
non-isothermal protoplanets it is nor possible to presently decide which peak
in the Mark~1 spectra will suffer the flattening, likely the more massive one
of the two main peaks but possibly both.

\section{Non-linear stability of short-period isothermal proto-planets}

Isothermal solution manifolds \citet{2005A&A...440.1183P} for short-period
(P=1..4d) proto-planets (PPs) have different non-linear stability properties
than those of long-period PPs (see \citet{2006Pecnik_Wuchterl} for dynamical
properties summary, and \citet{pecnik05} for $a = 5\,  {\rm AU}$ orbital region
details).

Unlike long-period PPs, the critical core mass ($M_\mathrm{crit}$) for
short-period isothermal PPs gets larger for smaller orbital periods. Isothermal
Hot-Jupiter's $M_\mathrm{crit}$ also strongly depends on the temperature and
the density of the surrounding nebula. Additionally, massive subcritical
short-period PPs are more linearly stable than long-period counterparts
\citet{2005DiplSchoenke}.

Fluid-dynamical calculations at orbital periods of 1 and 4d for a solar mass
primary, show that the non-linear stability properties of isothermal
subcritical PPs with temperatures of $T_\mathrm{1}$=1260 K and
$T_\mathrm{2}$=1600 K are almost identical, which hints that most of these
equilibria might be thermally stable. Short-period subcritical PPs with gas
density at the \underline{c}ore \underline{s}urface ($\varrho_\mathrm{cs}$) of
up to around 10 kg m$^{-3}$ are stable and just oscillate around the
equilibrium if perturbed. Subcritical PPs with higher $\varrho_\mathrm{cs}$
will make a transition towards a corresponding solution with lower
$\varrho_\mathrm{cs}$, and could either loose or gain envelope mass in the
process. Solution with still higher $\varrho_\mathrm{cs}$ will eject more than
95\% of its envelope. Finally, a compact PP (a PP with highest
$\varrho_\mathrm{cs}$) will remain stable, but will include an outgoing wind,
supersonic at the outer boundary. Depending on the level of envelope
compactness, mass loss due to this wind could be anything from completely
negligible to a significant percent of the envelope mass within a single
sound-crossing time. Additional effect of this wind is "puffing-up" of the
outer PP's stratifications. All of these properties suggest that massive stable
subcritical PPs are formed more easily at small orbital distances (than at
larger orbital distances), and that quasi-hydrostatic evolutionary paths are
available up to the $M_\mathrm{crit}$.

\section{Quasi-hydrostatic calculations and secular/thermal stability}

The previous sections were dedicated to the isothermal case that is
computationally very efficient but cannot provide quantitative agreement with
present day giant planet masses. To analyze the stability of the Mark~1
population directly, i.e. with the time-dependent versions corresponding to the
the static Mark~1 equations, a full stellar evolution type calculation is
necessary. Such calculations, with constitutive relations identical to the
Mark~1, have been performed for selected parts of the planetary manifolds for
orbital periods of 1 and 4 days for the solar mass case. The result is that in
all cases calculated, i.e.\ for core masses to $10^{23}\, {\rm kg}$ and total
masses up to about 50 earth masses the protoplanets contract and loose less
than 30\% of their total mass within the 100 Ma calculation period.

\section{Non-linear convective radiation-fluid-dynamic calculations}

Finally we performed full nonlinear-radiation fluid calculations for selected
protoplanetary models in continuation of \citet{2006tafp.conf...70B}. These
calculations are the largest computational effort and have been done in parts
of the CoRoT-Mark~1 planet population that are thought to be particularly
unstable from both the population analysis, \citet{2006DissBroeg} which finds
the planets at largest orbital distance to more likely  become dynamical and
the isothermal linear and non-linear studies,
\citet{2005DiplSchoenke,pecnik05}, that find the high envelope mass, high
nebula pressure, low core mass protoplanets to be the most likely ones to
suffer an instability. Because we are primarily interested in the stability
here, unlike \citet{2006tafp.conf...70B} we use a constant nebula state for the
outer boundary condition --- in practice that means an unlimited mass reservoir
instead of a feeding zone with given mass.

We used the equations of \citet{2003A&A...398.1081W} with planetary boundary
conditions, \citet{1991Icar...91...39W}. The dynamic evolution of 6 planets of
the Mark~1 population for a 16d period at $1\,{\rm M_{\rm \sun}}$ and core
masses of $0.47$ and $1.11\,{\rm M_{\rm earth}}$ was calculated with static
models as initial condition. These core masses are relatively low compared to
the typical critical masses for close-in planets, see \citet{2006DissBroeg}
to select potentially unstable planets. It turned out that the envelope mass
average of the first adiabatic exponent, $\Gamma_1$, an indicator for dynamical
stability in a stellar structure type context, has a characteristic depletion
below the critical value of $4/3$. That occurs as the `island' of large
planetary envelope masses (cf.\ \citet{2006DissBroeg}) is crossed along
constant core mass.

In all six cases
the evolution remains quasi-hydrostatic and the masses within 10\% of the
initial, static value for 2000a. Dynamically this means that \emph{all} the
respective protoplanets are stable. That in spite of selecting two models by
looking for the lowest $\Gamma_1$ at given core-mass. There is also no
indication of relevant mass loss due to a secular instability. The two most
massive planets with 105 and 126 earth masses, initially show a few percent
reduction in mass as the initially static planet starts to contract. After a
thermal adjustment of a $10^4 {\rm a}$ the mass levels off at a few percent
less than the initial values and starts growing.

\section{Close-in Convective Protoplanets}

The fact that `realistic´ close-in exoplanets turn out to be more stable than
estimated by isothermal models can be explained the fact that the envelopes are
largely convective. Largely convective protoplanets,
\citet{1993Icar..106..323W,2001ApJ...553..999I} damp dynamical instabilities in
their large convection zones, \citet{1995EM&P...67...51W} and are favour
accretion, \citet{2000prpl.conf.1081W}.

Thus as far as the sample calculations are representative of the ensemble, the
planets of the CoRoT~Mark~1 population are stable. The onset of apparent
adiabatic dynamical instability in 2/6 models is damped by convection.

\section{Discussion}

The CoRoT-Mark~1 theoretical planet populations rely on the implicit assumption
that there is no preferred scale in the masses of planetary cores and no
preferred range of nebula pressures. Together this requires to assume a
diversity of gravitationally stable nebulae. That does not imply arbitrary
ranges for core-masses and nebula pressures but sufficiently wide ranges so
that the planetary core-envelope equilibria can `draw' any required value from
a nebula at a given orbital radius. If some cores that are required in the
population do not form, the respective planets would not be able to form and
thus not appear in the observed mass spectra. Presently we do neither have a
sufficiently comprehensive overview about the diversity of nebula properties
from observations nor from theory. Thus the CoRoT-Mark~1 predictions rely on an
assumption of nebula diversity. Presently the best argument that there is
indeed a huge diversity of nebular conditions comes from the observation of
extreme planets as GQ Lupi b and HD 149 026 b.

The second key assumption to derive the mass spectra is that the planets are
static. While planet formation has been found to be a hydrostatic process under
many conditions and for most of the time, there are important exceptions and
there is no guarantee that the results from studies of planet formation for a
rather restricted set of conditions can be generalised to the diverse nebula
properties required by the Mark~1 population.

A general study of stability for the entire ensemble of $\sim 10^6$ `realistic'
protoplanets is presently beyond our computational reach. We therefore followed
a two-fold strategy. We studied simplified problems (linear isothermal
stability) in a comprehensive way and sampled a hierarchy of more detailed
models - nonlinear dynamical stability of isothermal PP, nonlinear secular
stability of realistic PP and did a limited set of non-linear radiation
fluid-dynamical calculations at positions that were indicated as particularly
unstable in the more comprehensive but simpler models. While the simpler models
provide a detailed and synoptic view, they indicate instabilities in some parts
of the planet-manifolds. Approximately 1/3 to 1/6, say, of the more massive
envelopes with small core are unstable and likely corresponding to realistic
Mark~1 planets. The few detailed models show that improved physics, in
particular time-dependent convection acts stabilising at the investigated
places an indicates stability for all cases calculated. In particular a
non-linear development of a case with linear adiabatic dynamic instability
indicated by globally low first adiabatic exponents (down to 1.25) is damped
out by convection and the respective planets remain static and thus stable,
valid elements of the Mark~1 planet population.

\section{Conclusions: CoRoT's planets}

Based on the assumption that diverse gravitationally stable protoplanetary
nebulae are produced by the star formation process we predict the short period,
$T_{\rm orb} < 16{\rm d}$ planet population of a typical CoRoT field:
    (1) they closely follow the CoRoT-Mark~1 planet population applied to the
        stellar population mix of the field, i.e.\ show planetary initial mass
        functions (IMFs) as given in Fig.~\ref{Mark1_CoRoT_Mix};
    (2) these IMFs have to be corrected for particle loss
        process to arrive at the epoch of observation mass function; this
        corrections are discussed in a parallel letter;
    (3) typical field planetary IMFs depend on orbital period and have the largest
        mass peak around a Jupiter mass, we refer to \citet{2006DissBroeg}
        for more predictions of properties of single-host-star-mass planetary IMFs.
    (4) the broad peaks in the stellar population mix IMFs
        point to a large population with masses of $1-3\,10^{26} {\rm kg}$, (50 --160
        earth-masses).
    (5) there is an abundant population of Hot Neptunes and an important
        population filling the solar-system mass gap between Neptune and
        Saturn.

\begin{acknowledgements}

The authors thank DLR and the german BMFT for support under grants DLR/MPE,
DLR/AIU, DLR /TLS. Bojan Pe\v{c}nik thanks the \emph{National Foundation for
Science, Higher Education and Technological Development} of \emph{the Republic
of Croatia}. G. Wuchterl thanks the MPE for computation-resources on DEC/VMS
machines.

\end{acknowledgements}

\bibliographystyle{aa}
\bibliography{Wuchterl,Wuchterl_ADS_2004}

%
%
%
%
%

\end{document}